\documentclass{article}
\usepackage{spconf,amsmath,graphicx,cite,bm,amsfonts}
\usepackage{resizegather,booktabs,multirow}

\newcommand{\argmax}[1]{\underset{#1}{\operatorname{max}}\;}
\newcommand{\argmin}[1]{\underset{#1}{\operatorname{min}}\;}
\newcommand{\norm}[1]{\left\lVert#1\right\rVert}
\title{Transformation of low-quality device-recorded speech to high-quality speech using improved SEGAN model}
\makeatletter
\def\@name{ Seyyed Saeed Sarfjoo$^1$, Xin Wang$^2$, Gustav Eje Henter$^2$, \\ Jaime Lorenzo-Trueba$^2$, Shinji Takaki$^2$, Junichi Yamagishi$^{2,3}$ 
\sthanks{This study was conducted during an internship of the first author at NII, Japan in 2017. This work was partially supported by MEXT KAKENHI Grant Numbers (15H01686, 16H06302, 17H04687).}}
\makeatother
\address{
  $^1$\"{O}zye\u{g}in University, Turkey, 
  $^2$National Institute of Informatics, Japan, 
  $^3$University of Edinburgh, UK \\
  {\small \tt saeed.sarfjoo@ozu.edu.tr, \{wangxin,gustav,jaime,takaki,jyamagis\}@nii.ac.jp}
  }
\begin{document}
\ninept
\maketitle
\begin{abstract}
Nowadays vast amounts of speech data are recorded from low-quality recorder devices such as smartphones, tablets, laptops, and medium-quality microphones. The objective of this research was to study the automatic generation of high-quality speech from such low-quality device-recorded speech, which could then be applied to many speech-generation tasks. In this paper, we first introduce our new device-recorded speech dataset then propose an improved end-to-end method for automatically transforming the low-quality device-recorded speech into professional high-quality speech. Our method is an extension of a generative adversarial network (GAN)-based speech enhancement model called speech enhancement GAN (SEGAN), and we present two modifications to make model training more robust and stable. Finally, from a large-scale listening test, we show that our method can significantly enhance the quality of device-recorded speech signals.
\end{abstract}
\begin{keywords}
Audio transformation, speech enhancement, generative adversarial network, speech synthesis
\end{keywords}
\section{Introduction}
\label{sec:intro}

Using high-quality speech recordings is essential in various speech-generation tasks such as speech synthesis and voice conversion. Currently, a large amount of speech-content sources, such as YouTube, podcasts, lecture videos, and audio stories, is available on the Web. Typically, such content is recorded in non-professional acoustic environments such as homes and offices. Moreover, the recordings are often carried out using consumer devices such as smartphones, tablets, and laptops. Therefore, the speech recordings of the content are of typically poor quality and contain a large amount of ambient noise and room reverberation. Even if the recordings are done under quiet conditions, they may still present low-quality standards due to using recording hardware with bad frequency characteristics and/or inappropriate bandwidth settings. In real applications, such as speaker adaptation of speech synthesis or voice conversion, we have to handle such non-ideal data in the wild; thus, we have to generate high-quality speech outputs. This objective may sound contradictory; however, there is a strong demand to achieve it. In this paper, we call this low-quality speech recorded using consumer devices ``device-recorded speech''.

One possible solution is the transformation from device-recorded speech to high-quality speech before voice conversion occurs or before the speech-synthesis models are trained, and this may be approached from two different directions \cite{mysore2015can}. One direction of handling device-recorded speech is to apply speech-enhancement techniques for denoising \cite{ephraim1984speech,scalart1996speech,duan2012speech}, dereverbration \cite{naylor2010speech,kinoshita2013reverb}, decoloration \cite{liang2014speech}, or bandwidth expansion \cite{enbom1999bandwidth}.
 
However, speech-enhancement techniques do not always handle quality degradation caused by hardware with bad frequency characteristics. We have to enhance clean but poor-quality speech recordings using hardware with bad frequency characteristics to high-fidelity speech. Therefore, the second direction is data-driven, non-linear, direct mapping from device-recorded speech to high-fidelity speech using machine-learning techniques such as deep learning \cite{bengio2013representation}.

In this paper, we propose a deep-learning-based method to transform low-quality device-recorded speech to high-quality speech. To that end, we recorded a new variant of the voice cloning toolkit (VCTK) dataset \cite{veaux2013voice}: device-recorded VCTK (DR-VCTK), where the high-quality speech signals recorded in a semi-anechoic chamber using professional audio devices are played back and re-recorded in office environments using relatively inexpensive consumer devices. Using the parallel database of the original VCTK and DR-VCTK, we can try the mapping between device-recorded and high-quality audio. Since the VCTK database includes a sufficient amount of speech data and we hypothesize that degradation due to frequency characteristics of microphones and loudness speakers, as well as degradation due to noise and reverberation, is beyond what is assumed with normal speech enhancement, we use deep-learning techniques for the new mapping problem instead of signal-processing techniques such as Wiener filtering \cite{lim1979enhancement}.

The chosen neural-network-mapping model  is the recently proposed speech enhancement generative adversarial network (SEGAN) \cite{pascual2017segan}, which is an end-to-end model for directly enhancing the noisy speech in the time domain. This is in contrast to previous DNN-based speech-enhancement techniques, which are based on short-time Fourier analysis/synthesis. Rethage's speech denoising model \cite{rethage2017wavenet} and WaveMedic \cite{fisherwavemedic} are other time domain-based speech enhancers that use the WaveNet model \cite{oord2016wavenet} for enhancing the degraded speech signal. However, unlike Wavenet, SEGAN is a regression model and consists of a convolutional network architecture trained using the GAN criterion \cite{goodfellow2014generative}. Using this time domain-based architecture, we can expect that the phase spectrum of the noisy signal may be transformed into the clean phase spectrum, and that it may have a good effect in improving speech quality \cite{paliwal2011importance}. The use of a GAN may also alleviate the over-smoothing problem and improve the quality of enhanced speech.

In our preliminary experiments, however, we found that even the recent SEGAN is negatively affected when mapping from device-recorded speech to high-quality speech recorded in a semi-anechoic chamber using professional devices. Therefore, we propose a new training procedure for the SEGAN model to improve the final quality of enhanced speech. The key of the proposed method is to use directed references for training SEGAN at the initial training epochs. By using this directed reference for training the generator model, we can achieve better weight initialization; thus, we were able to robustly and quickly train the generator model. We also found that this method  significantly reduces the appearance of annoying artifacts called ``musical noise'', something that conventional-speech-enhancement methods are commonly affected by. The performance of the proposed method  was evaluated through objective and subjective experiments.

This paper is structured as follows: We introduce the new DR-VCTK dataset in Section~\ref{sec:DB}. In Section~\ref{sec:SEGAN}, we describe SEGAN model for speech enhancement. We describe the proposed training procedure  of SEGAN in Section~\ref{sec:DIGAN} and the experimental setup and objective and subjective evaluation results in Section~\ref{sec:Exp}. Finally, we give conclusions and discuss future work in Section~\ref{sec:Conclusion}.

\vspace{-3mm}
\section{Device-recorded VCTK}
\label{sec:DB}

We used the centre for speech technology research (CSTR) VCTK corpus \cite{veaux2013voice} as the clean speech-signal source for the device-recorded signals, as it was recorded at high-quality using professional audio devices. This dataset contains recordings of 109 English speakers with different accents. There are around 400 sentences available from each speaker. 

\begin{figure}[tb]
	\centering
	\includegraphics[width=0.45\textwidth ,trim={0cm 4.3cm 20cm 12.0cm},clip]{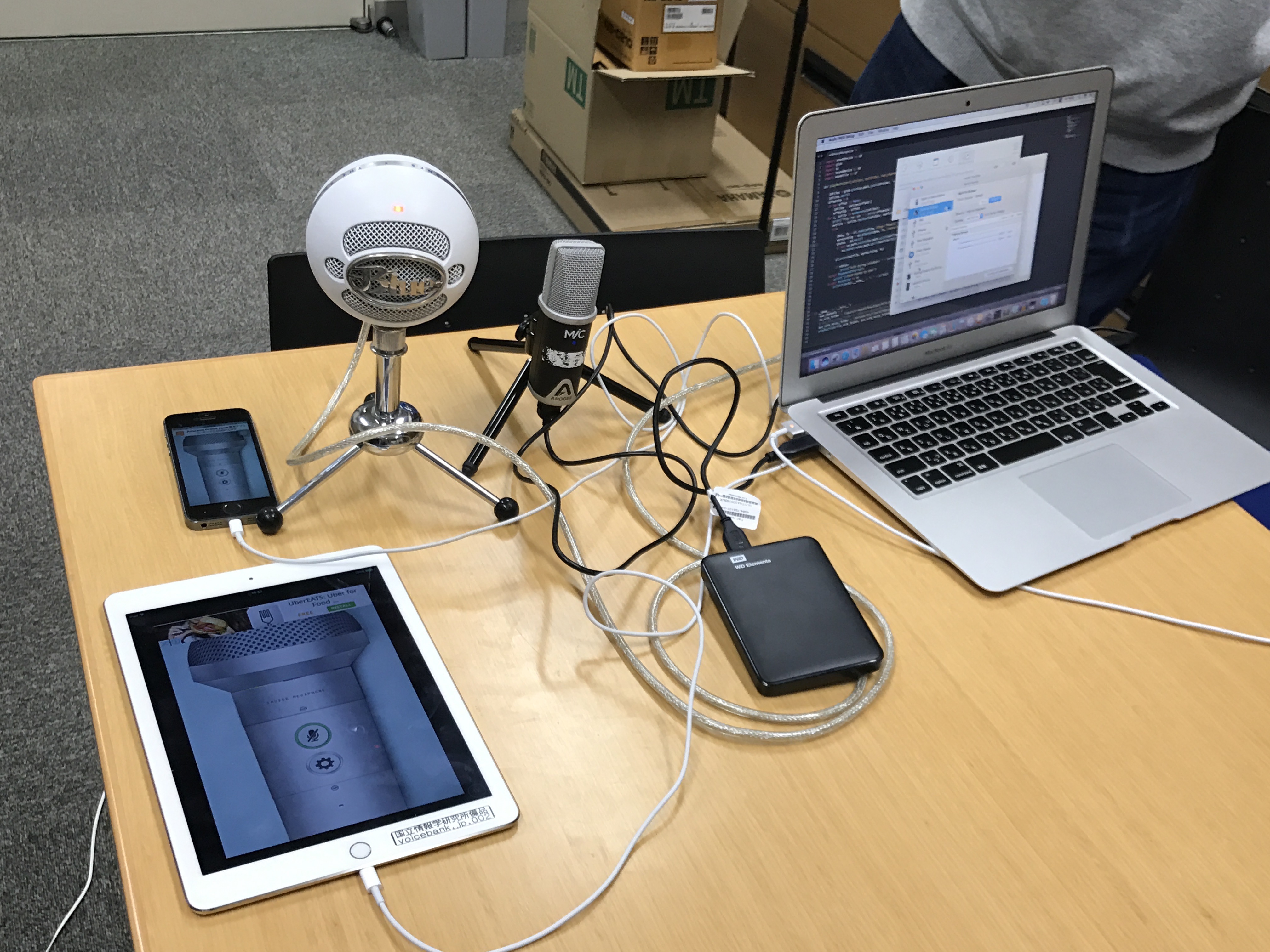}
	\caption{Setup for device-recording in office. The clean studio recordings were played through loudspeaker and recorded on either iPad, iPhone 5S, Mac Book Air, Blue Snowball, or Apogee Mic microphones. This setup captured noise and reverberation of room as well as limitations of recording hardware.}
	\label{fig:recsetup}
	\vspace{-5mm}
\end{figure}

Audio signals included in the CSTR VCTK corpus were played back from a loudspeaker and re-recorded using relatively inexpensive consumer devices in office environments. We used eight different microphones for the recording of device-recorded speech signals (MacBookAir's two microhpones, Apogee MiC, Blue Snowball, iPhone 5S's two microphones, and iPad's two microphones). The setup for device-recording is shown in Fig. \ref{fig:recsetup}. Bose 404600 SoundLink speaker III was used as a high-quality speaker and was set 2 meters from the microphones. Recording was done in a medium-sized office under two background-noise conditions (i.e. windows either opened or closed). We recorded device-recorded signals under 16 conditions (8 microphones x 2 background noise conditions). All data were sampled at 48 kHz.

Among the 109 speakers, we selected 28 speakers (14 male and 14 female with British received pronunciation accent) for training and selected 2 speakers (1 male and 1 female) who had the same accent for testing. Twelve out of the 16 recording conditions were used for training and the remaining 4 recording conditions were used for testing. Half of the recording conditions in the training set (6 out of 12 sets) and half of those in the test sets (2 out of 4 sets) were selected from the windows-open background-noise condition. In other words, there was neither overlapped speakers nor recording conditions between training and test sets. However, each of the training and test sets included speech data under both windows-open and windows-closed background-noise conditions.

We used auto-correlation for removing the delay between clean and playback data. Silence segments longer than 200 ms were trimmed from the beginning and end of each sentence. To have a suitable input-chunk size in training, we down-sampled the dataset to 16 kHz.\footnote{This processed subset \cite{sarfjoo2018device} is publicly available from https://doi.org/10.7488/ds/2316.}

We also used a publicly available noisy-speech dataset \cite{valentini2016speech} for fair comparison of our proposed method with previous methods under wider type  and noisy conditions. This dataset is a collection of artificially corrupted noisy speech based on the CSTR VCTK corpus, publicly available in the DataShare repository of University of Edinburgh\footnote{http://datashare.is.ed.ac.uk/handle/10283/1942}. We call this dataset the Edinburgh noisy speech dataset. Since both datasets are based on the CSTR VCTK corpus, speakers and utterances of the Edinburgh noisy speech dataset are similar to those of the DR-VCTK dataset presented above. For the training set, 40 different conditions were considered \cite{valentini2016speech}: 10 types of noise (2 artificial and 8 from the Demand database) with 4 signal-to-noise ratios (SNR) each (15, 10, 5, and 0 dB). There were around ten different sentences for each condition per training speaker. To make the test set, a total of 20 different conditions were considered \cite{valentini2016speech}: five types of noise (all from the Demand database) with four SNRs each (17.5, 12.5, 7.5, and 2.5 dB). There were around 20 different sentences for each condition per test speaker. For this experiment, we down-sampled the dataset to 16 kHz.


\vspace{-3mm}
\section{GAN-based Waveform Enhancement}
\label{sec:SEGAN}

Generative adversarial nets were introduced as a novel way to train a generative model. They consist of two ``adversarial'' models: a generative model $G$ that captures the data distribution and a discriminative model $D$ that estimates the probability that a sample came from the training data rather than $G$. To learn the generator distribution $p_g$ over data $\bm{x}$, the generator builds a mapping function from a prior noise distribution $p_z(\bm{z})$ to the data space as $G(\bm{z};\theta_g)$. The discriminator $D(\bm{x};\theta_d)$ outputs a single scalar representing the probability that $\bm{x}$ came from training data rather than $p_g$ \cite{goodfellow2014generative}.

The $G$ and $D$ are both trained simultaneously. The parameters for $G$ are adjusted to minimize $\log (1 - D (G (\bm{z})))$ and parameters for $D$ are adjusted to maximize $\log(D(\bm{x}))$, as if they were following the two-player min-max game with value function $V(D, G)$:
\begin{equation}
\label{eq:GAN1}
\begin{split}
\argmin{G}\argmax{D}V(D, G) = \mathbb{E}_{\bm{x} \sim p_{\textsf{data}}(\bm{x})}[\log(D(\bm{x}))]  +     \\ 
\mathbb{E}_{\bm{z} \sim p_{\textsf{z}}(\bm{z})}[\log(1-D(G(\bm{z})))].
\end{split}
\end{equation}
This model can be extended with a conditioned version of a GAN, where we have extra information in $G$ and $D$ to execute mapping and classification \cite{mirza2014conditional}. In this case, we added an extra input $\bm{x}_c$ from which we change the objective function to

\begin{equation}
\label{eq:CGAN}
\begin{split}
\argmin{G}\argmax{D}V(D, G) = \mathbb{E}_{\bm{x} \sim p_{\textsf{data}}(\bm{x},\bm{x}_c)}[\log(D(\bm{x}, \bm{x}_c))]  +     \\ 
\mathbb{E}_{\bm{x}_c \sim p_{\textsf{data}}(\bm{x}_c), \bm{z} \sim p_{\textsf{z}}(\bm{z})}[\log(1-D(G(\bm{z}, \bm{x}_c)))].
\end{split}
\end{equation}

The original GAN approach was affected by the vanishing gradient problem due to the sigmoid cross-entropy loss function that was used to compute the cost. To solve this, the least-squares GAN (LSGAN) approach was proposed \cite{mao2016least}, which substitutes the cost function by the least-squares function with binary coding (1 is real, 0 is fake). With this approach, the formulation in Eq.~\ref{eq:CGAN} changes to
\begin{align}
\label{eq:LSGAND}
\argmin{D}V'(D) =& \frac{1}{2}\mathbb{E}_{\bm{x} \sim p_{\textsf{data}}(\bm{x},\bm{x}_c)}[(D(\bm{x}, \bm{x}_c)-1)^2]  +     \\ 
& \frac{1}{2}\mathbb{E}_{\bm{x}_c \sim p_{\textsf{data}}(\bm{x}_c), \bm{z} \sim p_{\textsf{z}}(\bm{z})}[D(G(\bm{z}, \bm{x}_c))^2] \nonumber\\
\argmin{G}V' (G) =&  
\mathbb{E}_{\bm{x}_c \sim p_{\textsf{data}}(\bm{x}_c), \bm{z} \sim p_{\textsf{z}}(\bm{z})}[(D(G(\bm{z}, \bm{x}_c))-1)^2].
\label{eq:LSGANG}
\end{align}

Based on the criterion of LSGAN, SEGAN model \cite{pascual2017segan} has been proposed for the task of speech enhancement. 
The generator in SEGAN $G(.)$ adopts a decoder-encoder structure to convert the degraded speech waveform $\tilde{\bm{x}}$ and a random vector $\bm{z}$ into an enhanced waveform $\hat{\bm{x}}$, which can be written as $\hat{\bm{x}} = G(\tilde{\bm{x}}, \bm{z})$. Specifically, the encoder in $G(.)$ uses multiple strided convolution layers to transform $\tilde{\bm{x}}$ into an embedded vector $\bm{c}$. After concatenating $\bm{c}$ and $\bm{z}$, the decoder part uses several fractional-strided convolution layers to produce $\hat{\bm{x}}$. Note that skip connections are added to connect each encoding layer to its homologous decoding layer, which is expected to facilitate the feature propagation between the encoder and decoder. The discriminator in SEGAN $D(.)$ takes $\tilde{\bm{x}}$ or the clean natural waveform as the input then outputs a real-valued number that can be used to evaluate the least-square criterion in Eqs.~(\ref{eq:LSGAND}) and (\ref{eq:LSGANG}). The $D(.)$ has a similar convolutional structure as the encoder in $G(.)$, however, it includes an additional $1\times1$ convolution layer and a fully connected output layer with a linear activation function. Although SEGAN can be directly trained on the basis of LSGAN's criterion, it was found that SEGAN performed better when an the $L_1$ norm term was added to Eq.~(\ref{eq:LSGANG}), which becomes
\begin{equation}
\label{eq:LSGANG2}
\begin{split}
\argmin{G}V''(G) =& 
\mathbb{E}_{\bm{x}_c \sim p_{\textsf{data}}(\bm{x}_c), \bm{z} \sim p_{\textsf{z}}(\bm{z})}[(D(G(\bm{z}, \bm{x}_c))-1)^2]\\
& + \lambda\norm{G(\bm{z},\tilde{\bm{x}})-\bm{x}}_1.
\end{split}
\end{equation}
Here the $L_1$ norm is weighted by a hyper-parameter $\lambda$.

\section{Robust SEGAN training}
\label{sec:DIGAN}
\begin{figure}[tb]
	\centering
	\includegraphics[width=0.48\textwidth ,trim={0cm 5.3cm 1cm 4.0cm},clip]{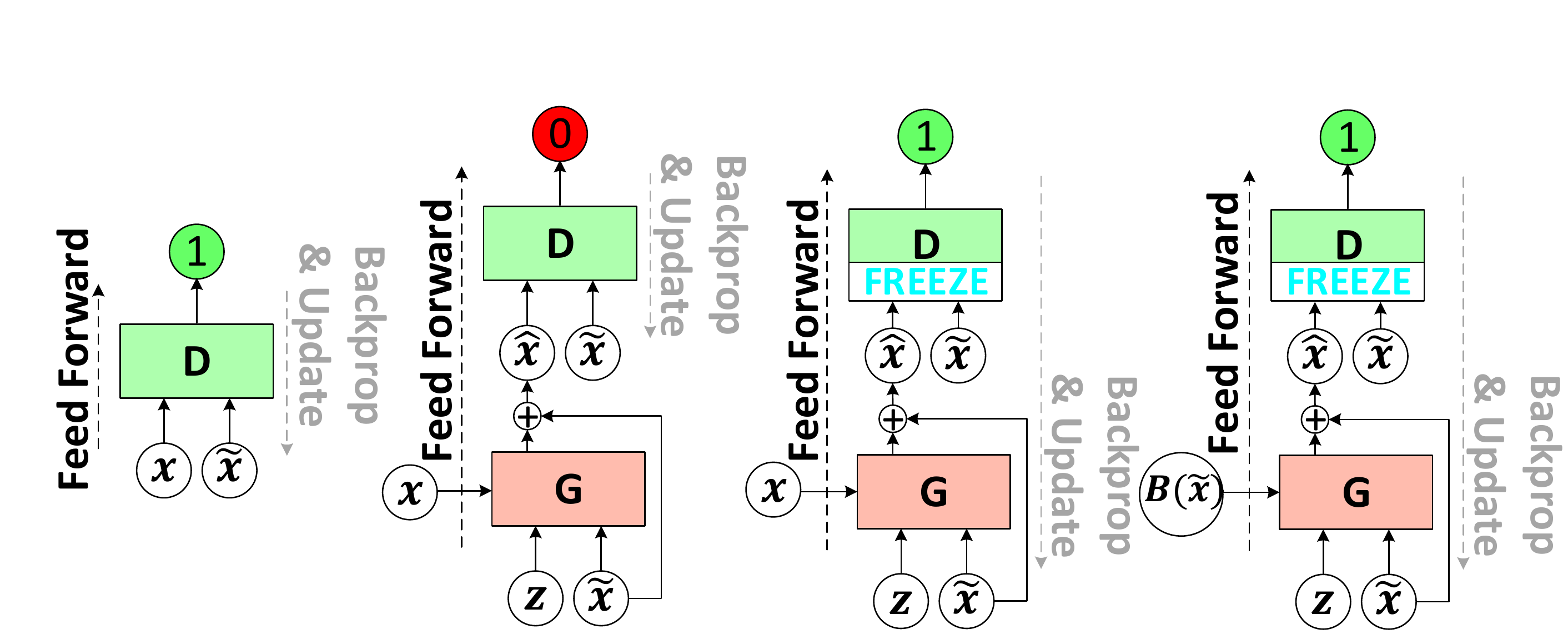}
	\caption{Steps of the modified SEGAN training strategy.}
	\label{fig:DIGANSteps}
	\vspace{-5mm}
\end{figure}

In this study, we started with SEGAN \cite{pascual2017segan}; however, it was found that the training process was sensitive to the noise levels in the input speech. To make the training process more robust and stable, we introduce a modified training strategy for the generator of SEGAN. Suppose a baseline speech-enhancement model $B(.)$, either a simple signal processing module or an unsophisticated neural network, is available for generating enhanced waveforms $B(\tilde{\bm{x}})$. The proposed strategy is to replace the clean signal $\bm{x}$ in Eq.~(\ref{eq:LSGANG2}) with $B(\tilde{\bm{x}})$ at the initial training phase. In this model, after K iterations in discriminator \cite{goodfellow2014generative}, we have J iterations in generator. Accordingly, Eq.~(\ref{eq:LSGANG2}) can be re-written as
\begin{align}
\label{eq:DIGANG}
\argmin{G}V'''_{i}(G) =& 
\mathbb{E}_{\bm{x}_c \sim p_{\textsf{data}}(\bm{x}_c), \bm{z} \sim p_{\textsf{z}}(\bm{z})}[(D(G(\bm{z}, \bm{x}_c))-1)^2]\\
& + \lambda\norm{G(\bm{z},\tilde{\bm{x}})-\bm{r}_i}_1, \nonumber
\end{align}
where $i$ is the index of J iterations for $G(.)$ and $\bm{r}_i$ is specified according to the schedule
\begin{gather}
\label{eq:DIGANG2}
\begin{split}
    \bm{r}_i= 
\begin{cases}
    B(\tilde{\bm{x}})& \text{if } 1-\frac{i}{J}\leq P_J,  ~~~~0\leq i<J\\
    \bm{x},              & \text{otherwise},
\end{cases}
\end{split}.
\end{gather}
In the above schedule, $P_J$ is the predefined probability for selecting $B(\tilde{\bm{x}})$ instead of $\bm{x}$. By using the above criterion, with $P_J$ probability, $B(\tilde{\bm{x}})$ rather than $\bm{x}$ is used for weight initialization in the initial training epochs. We hypothesize that SEGAN can be trained more stably as the pre-enhanced sample $B(\tilde{\bm{x}})$ is less stochastic than $\bm{x}$; however, it still sounds relatively clean. The modified SEGAN training strategy is illustrated in Fig. \ref{fig:DIGANSteps}.

For further improving SEGAN training, an additional skip-connection  was added around the generator $G(.)$. Differing from the original skip-connections between the encoder and decoder in $G(.)$, the proposed skip-connection directly delivers the input $\tilde{\bm{x}}$ to the output side of $G(.)$. Accordingly, the generated enhanced speech becomes $\hat{\bm{x}} = G(\tilde{\bm{x}}, \bm{z})+\tilde{\bm{x}}$. In this way, the task of $G(\tilde{\bm{x}}, \bm{z})$ is not to generate enhanced speech from scratch but to generate a residual signal that refines the input speech \cite{kaneko2017generative}. By replacing $G(\tilde{\bm{x}}, \bm{z})$ in Eq.~\ref{eq:DIGANG} with $G(\tilde{\bm{x}}, \bm{z})+\tilde{\bm{x}}$, the proposed method encourages the generator to learn the detailed differences between clean and enhanced speech waveforms. 


\section{Experiments and Results}
\label{sec:Exp}
Similar to the original SEGAN training strategy, we extracted chunks of waveforms with a sliding window of $2^{14}$ samples at every $2^{13}$ samples (i.e. 50\% overlap). At testing time, we concatenated the results at the end of the stream without overlapping. For the last chunk, instead of zero padding, we pre-padded it with the previous samples. For batch optimization, RMSprop \cite{tieleman2012lecture} with 0.0002 learning rate and batch size of 100 was used. The modified SEGAN model converged at 120 epochs, and we set $J$ to 2 and $P_J$ to 50\% for the first 50 epochs.

For selecting the pre-enhancement method, we compared Wiener \cite{lim1979enhancement}, harmonic regeneration noise reduction (HRNR) \cite{plapous2006improved}, and Postfish \cite{postfish2005} algorithms. In our preliminary experiments, applying Postfish and HRNR sequentially showed better quality enhanced samples. We used this compound method to generate $B(\tilde{\bm{x}})$ in Eq.~\ref{eq:DIGANG2}.

Based on our initial experiments, we fixed the $\lambda$ in Eq.~\ref{eq:DIGANG} to 100. With this configuration, we observed equilibrium behavior in the adversarial training and obtained samples with better quality. Like the generator in original SEGAN, we used 22 one-dimensional strided convolution layers with a filter width of 31 and stride of 2. The encoder part of the generator used 11 strided convolution layers. If the size of each layer's output feature matrix is denoted by length $\times$dimension, then this size changes as 8192$\times$16, 4096$\times$32, 2048$\times$32, 1024$\times$64, 512$\times$64, 256$\times$128, 128$\times$128, 64$\times$256, 32$\times$256, 16$\times$512, and 8$\times$1024. The output of the encoder is then concatenated with the latent vector $\bm{z}$, which was drawn from a normal distribution of dimension 8$\times$1024. The decoder part was a mirror of the encoder part, except for the additional skip connections and input latent vector, which doubled the number of feature maps in every layer.

The discriminator network is like the encoder part of the generator network; however, it uses virtual batch-norm \cite{salimans2017improved} before LeakyReLU non-linearities with $\alpha = 0.3$ followed by $1\times1$ convolution and one fully connected layer with a linear activation function. The implementation of the improved SEGAN is publicly available\footnote{https://github.com/ssarfjoo/improvedsegan}.

\vspace{-2mm}
\subsection{Objective Evaluation}
\label{subsec:ObjectiveEval}

We first objectively compared the performance of SEGAN and other baseline models with that of the proposed model. Even if transformation of low-quality device-recorded speech to high-quality speech is a different task from conventional speech enhancement, the objective measures are still relevant. Therefore, we discuss the objective measures used for speech enhancement in addition to other objective measures. Using the following objective measures (the higher the better), this evaluation was done on both DR-VCTK and Edinburgh datasets:

\begin{itemize}
    \setlength{\itemsep}{0mm}
    \item CSIG: Mean opinion score (MOS) prediction of the signal
distortion attending only to the speech signal (from 1
to 5).
    \item CBAK: MOS prediction of the intrusiveness of background
noise (from 1 to 5).
    \item COVL: MOS prediction of the overall effect (from 1
to 5).
    \item PESQ: Perceptual evaluation of speech quality, a
metric used in telecommunications for estimating the
perceived quality of speech audio. Five-point scale
in MOS-LQO (from -0.5 to 4.5).
    \item SSNR: Segmental SNR (from -10dB to 35dB). See \cite{papamichalis1987practical}.
    \item DAU: Prediction
of speech intelligibility based on an auditory pre-processing
model (from 0 to 1).
  \item STOI: An algorithm for intelligibility predicting time-frequency weighted noisy speech (from 0 to 1).
\end{itemize}

\renewcommand{\arraystretch}{1.2}
\begin{table}[tb]
\centering
\caption{Objective evaluation on DR-VCTK and Edinburgh datasets}
\label{table:Objective}
\resizebox{\linewidth}{!}{%
\begin{tabular}{lc|c|c|c|c|c|c|c|}
\cline{3-9}
 & \multicolumn{1}{l|}{} & \multicolumn{1}{c|}{\textbf{CSIG}} & \multicolumn{1}{c|}{\textbf{CBAK}} & \multicolumn{1}{c|}{\textbf{COVL}} & \multicolumn{1}{c|}{\textbf{PESQ}} & \multicolumn{1}{c|}{\textbf{SSNR}} & \multicolumn{1}{c|}{\textbf{DAU}} & \multicolumn{1}{c|}{\textbf{STOI}} \\ \hline
\multicolumn{1}{|l|}{\parbox[t]{2mm}{\multirow{4}{*}{\rotatebox[origin=c]{90}{\textbf{DR-VCTK}}}}} & \textbf{Noisy} & \textbf{2.17} & 1.43 & \textbf{1.58} & 1.24 & -3.66 & 0.71 & 0.72 \\ \cline{2-9} 
\multicolumn{1}{|l|}{} & \textbf{Postfish+HRNR} & 1.62 & \textbf{1.63} & 1.31 & 1.27 & -1.66 & 0.69 & 0.72 \\ \cline{2-9} 
\multicolumn{1}{|l|}{} & \textbf{Original SEGAN} & 1.66 & 1.60 & 1.32 & 1.24 & \textbf{-1.09} & 0.58 & 0.65 \\ \cline{2-9} 
\multicolumn{1}{|l|}{} & \textbf{Proposed SEGAN} & 1.96 & 1.60 & 1.50 & \textbf{1.28} & -1.72 & \textbf{0.72} & \textbf{0.73} \\ \hline\hline
\multicolumn{1}{|l|}{\parbox[t]{2mm}{\multirow{4}{*}{\rotatebox[origin=c]{90}{\textbf{Edinburgh}}}}} & \textbf{Noisy} & \textbf{3.34} & 2.44 & \textbf{2.63} & 1.97 & 1.73 & 0.90 & 0.92 \\ \cline{2-9} 
\multicolumn{1}{|l|}{} & \textbf{Postfish+HRNR} & 2.11 & 2.38 & 1.95 & 1.93 & 6.26 & 0.87 & 0.90 \\ \cline{2-9} 
\multicolumn{1}{|l|}{} & \textbf{Original SEGAN} & 3.00 & \textbf{2.65} & 2.55 & \textbf{2.14} & \textbf{8.21} & \textbf{0.92} & \textbf{0.93} \\ \cline{2-9} 
\multicolumn{1}{|l|}{} & \textbf{Proposed SEGAN} & 2.32 & 2.49 & 2.07 & 1.94 & 6.33 & 0.89 & 0.91 \\ \hline
\end{tabular}%
}
\end{table}
\renewcommand{\arraystretch}{1.0}

The evaluation results are listed in Table \ref{table:Objective}. In the DR-VCTK dataset, quality (PESQ) and intelligibility (DAU and STOI) measures of the improved SEGAN model were better than those of the original SEGAN and a combination of Postfish and HRNR. On the other hand, in the Edinburgh dataset, in which the energy of noise is clearly lower than the DR-VCTK dataset, the original SEGAN had better scores than the other models in terms of PESQ, SSNR, DAU, and STOI.

\subsection{Subjective Evaluation}
\label{subsec:SubjectiveEval}

We also carried out a crowdsourced subjective evaluation to rate the end-user impact of the proposed enhancement models . The evaluation was aimed to rate the subjective listener's perception in a 1-to-5 MOS framework, as is commonly done in speech-synthesis evaluations \cite{bc2011}. To make the question clearer to the evaluators, we modified the wording to explicitly ask them to rate each sample in terms of the degree of noise and sound-quality degradation, ranging from 1 (noise and/or quality degradation are clearly audible) to 5 (there is no speech-quality degradation or noise).

The evaluation was carried by means of a web interface, where the listeners were presented with a set of two blocks, each consisting of nine screens. Each block contained one screen for each of the nine evaluated conditions (i.e.,\ 4 models  $\times$ 2 datasets plus the clean speech reference), always using the same evaluation utterance, randomly selected from the test-sentence pool. That is, all nine conditions were rated before listening to them again in a second different utterance. The ordering of the systems in each block was also randomized. Each screen then contained one evaluation sample and one evaluation question. The samples could be played as many times as desired by the evaluators, and they were not allowed to proceed to the next system until the current one was played to completion and rated. A total of 107 native Japanese speakers took part in the evaluation, for a total of 1236 sets, i.e., 22248 evaluation samples or 2472 per condition. The evaluation results are listed in Table~\ref{table:subjectiveEvaluation}.


\begin{table}[tb]
\centering
\caption{Subjective evaluation results on DR-VCTK and Edinburgh datasets. MOS score for clean speech was $4.34$.}
\label{table:subjectiveEvaluation}
\resizebox{\linewidth}{!}{
\begin{tabular}{c|c|c|c|c|}
\cline{2-5}
\multicolumn{1}{l|}{}                    &                & \textbf{Postfish} & \multicolumn{2}{c|}{\textbf{SEGAN}}  \\ \cline{4-5}
\multicolumn{1}{l|}{}                    & \textbf{Noisy} & \textbf{+HRNR} & \textbf{Original}  & \textbf{Proposed } \\ \hline
\multicolumn{1}{|c|}{\textbf{DR-VCTK}}   & $2.54$  & $2.78$ & $1.14$  & $\textbf{2.80}$           \\ \hline
\multicolumn{1}{|c|}{\textbf{Edinburgh}} & $2.84$  & $3.29$  & $3.40$  & $\textbf{3.44}$  \\ \hline
\end{tabular}
}
\vspace{-5mm}
\end{table}

To study the significance of the subjective results, we carried out unpaired t-test comparisons for a 95\% confidence with Holm-Bonferroni compensation to take into account the multiple pairwise comparisons. The analysis showed that our proposed SEGAN method  is comparable to a combination method of postfish and HRNR on the DR-VCTK dataset ($p$-value = 0.39691); however, both were significantly better than the original SEGAN ($p$-value $< 2e^{-16}$). Also, the proposed SEGAN was comparable to the original SEGAN in the Edinburgh dataset ($p$-value = 0.39691); however, it was significantly better than the combination method of postfish and HRNR ($p$-value = 0.00011). This means that the proposed method is more noise-robust and stable than the original version. We also hypothesize that the contrast between the objective measures and subjective results might be due to the SEGAN samples not having musical noise artifacts.


\vspace{-2mm}
\section{Conclusions and Future Work}
\label{sec:Conclusion}
We proposed a method for automatically transforming low-quality device-recorded speech to high-quality speech. To that end, we recorded the low-quality device-recorded version of the VCTK corpus, DR-VCTK. We also proposed an improved SEGAN training algorithm  using pre-enhanced samples instead of clean data as ground truth data. From a large-scale listening test, we confirmed that our method can enhance the perceptual quality of speech signals on both DR-VCTK and Edinburgh datasets. Our future work includes studying SEGAN-based transformation of low-quality sounds such as device-recorded musical sounds.


\bibliographystyle{IEEEbib}
\bibliography{strings,refs}

\end{document}